\documentclass[12pt]{iopart}
\usepackage{graphicx}

\begin{document}

\title[]{Multiplicity fluctuations 
       in relativistic nuclear collisions: 
       statistical model versus experimental data}

\author{M Hauer$^1$, V V Begun$^2$, M Ga\'zdzicki$^{3,4}$, M I Gorenstein$^{2,5}$,\\ V P
  Konchakovski$^{1,2}$ and  B Lungwitz$^4$}

\address{$^1$ Helmholtz Research School, University of Frankfurt, Frankfurt,
  Germany}
\address{$^2$ Bogolyubov Institute for Theoretical Physics, Kiev, Ukraine}
\address{$^3$ Institut f\"ur Kernphysik, University of Frankfurt, Frankfurt,
  Germany}
\address{$^4$ \'Swi\c{e}tokrzyska Academy, Kielce, Poland}
\address{$^5$ Frankfurt Institute for Advanced Studies, Frankfurt,Germany}

\ead{hauer@fias.uni-frankfurt.de}

\begin{abstract}
Multiplicity distributions of hadrons produced in central nucleus-nucleus
collisions are studied within the hadron-resonance gas model in the large
volume limit. In the canonical ensemble conservation of three charges
(baryon number, electric charge, and strangeness) is enforced. In addition, in 
the micro-canonical ensemble energy conservation is included. An analytical
method is used to account for resonance decays. Multiplicity distributions
and scaled variances for negatively charged hadrons are presented along
the chemical freeze-out line of central Pb+Pb (Au+Au) collisions from SIS to
LHC  energies. Predictions obtained within different statistical ensembles are
compared with preliminary NA49 experimental results on central Pb+Pb
collisions in the SPS energy range. The measured fluctuations are
significantly narrower than a Poisson reference distribution, and clearly favor
expectations for  the micro-canonical ensemble.  

\end{abstract}

\maketitle

\section{Introduction}
\label{Intro}
For more than 50 years statistical models of strong interactions
\cite{fermi,landau,hagedorn} have served as an important tool to
investigate high energy nuclear collisions. The main subject of
the past study has been the mean multiplicity of produced hadrons (see
e.g. Refs. \cite{stat1,pbm,FOC,FOP}). Only recently first measurements of
fluctuations of particle multiplicity \cite{fluc-mult} and transverse momenta
\cite{fluc-pT} were performed. The
growing interest in the study of fluctuations in strong interactions is
motivated by expectations of anomalies in the vicinity of the onset of
deconfinement \cite{ood} and in the case when the expanding system goes
through the transition line between quark-gluon plasma and hadron gas
\cite{fluc2}. In particular, a critical point of strongly interacting matter
may be signaled by a characteristic power-law pattern in fluctuations
\cite{fluc3}. Apart from being an important tool in an effort to study the
critical behavior, the study of fluctuations within the statistical
hadronization model constitutes an essential test of its validity. 

Fluctuations are quantified by the ratio of variance of a
multiplicity distribution to its mean value, the so-called scaled
variance. There is a qualitative difference in the properties of mean
multiplicity and scaled variance of multiplicity distributions
in statistical models. In the case of mean multiplicity
results obtained within the grand canonical ensemble (GCE), canonical
ensemble (CE), and micro-canonical ensemble (MCE)  approach  each
other in the large volume limit. One refers here to as the
thermodynamical equivalence of statistical ensembles. It was
recently pointed out \cite{CE}, that corresponding results for the
scaled variance are different in different ensembles, and thus the
scaled variance is sensitive to conservation laws obeyed by a
statistical system. The differences are preserved  in the
thermodynamic limit.

In this contribution we briefly summarize recent results \cite{mce} on
multiplicty fluctuations in the hadron resonance gas model.

We will first discuss a simple example, accessible to analytical solutions,
and then sketch how to generalize the procedure to a general multi-specie
hadron gas. Lastly a comparison of model calculations to recently released
NA49 data \cite{NA49} on charged particle multiplicity fluctuations is shown.

\section{Ultra-relativistic Gas of Neutral Particles}
The GCE partition function of an ultra-relativistic gas composed of only
neutral Boltzmann particles is given by:
\begin{equation}
Z^{GCE}(V,T)=\exp \left[ V g \int \frac{d^3p}{\left(2\pi \right)^3}
        e^{-|p|/T} \right] = \exp \left[ Vg \frac{T^3}{\pi^2}\right],
\end{equation}
where $V$ is the volume of the system, $T$ its temperature, and $g$ the
degeneracy factor due to the particles internal spin. The number of
GCE micro-states with fixed particle number $N$ is given by the Fourier
integral over the generalized GCE partition function:
\begin{equation}
\mathcal{Z}^N(V,T) = \int \limits_{-\pi}^{\pi}
    \frac{d\phi_N}{2\pi}~e^{-iN\phi_N}~ \exp 
    \left[ Vg \frac{T^3}{\pi^2} e^{i\phi_N}\right] = \frac{\left(Vg
        \frac{T^3}{\pi^2}\right)^N}{N!}, 
\end{equation}
where the Wick-rotated fugacity $e^{i\phi_N}$ is introduced to fix
particle number $N$. The normalized multiplicity distribution, i.e. the
probability to find the system in a state with exactly $N$ particles is then
given by the ratio: 
\begin{equation}\label{P_gce}
P_{GCE}(N)~\equiv~ \frac{\mathcal{Z}^N(V,T)}{Z^{GCE}(V,T)}  ~=~ \frac{\left(Vg
    \frac{T^3}{\pi^2}\right)^N}{N!}~\exp \left(- Vg \frac{T^3}{\pi^2}\right)~.
\end{equation}
Similarly one can find the number of micro-states in GCE with fixed
particle number $N$ and fixed energy $E$:
\begin{eqnarray}
\fl \mathcal{Z}^{N,E}(V,T)&=& \int \limits_{-\pi}^{\pi} \frac{d\phi_N}{2\pi}
    \int \limits_{-\infty}^{\infty} \frac{d\phi_E}{2\pi}~ e^{-iN\phi_N}~
    e^{-iE\phi_E} \exp \left[ V g \int  \frac{d^3p}{\left(2\pi \right)^3}
      e^{-|p|/T}  e^{i |p|\phi_E} e^{i\phi_N}\right] \nonumber \\
    &=& \left(\frac{gV}{\pi^2} \right)^N ~\frac{E^{3N-1}}{N!\left(3N-1 
      \right)!} ~ e^{-E/T}~=~ Z^{MCE} (V,E,N)~ e^{-E/T}~, \label{massless_int}
\end{eqnarray}
where $ Z^{MCE} (V,E,N)$ denotes the MCE partition function. The number of
micro-states with fixed energy $E$, but arbitrary particle 
number $N$ is simply $\mathcal{Z}^{E}(V,T) = \sum \limits_{N=1}^{\infty}
\mathcal{Z}^{E,N}(V,T)$ in GCE, or $Z^{MCE}(V,E) = \sum \limits_{N=1}^{\infty}
Z^{MCE}(V,E,N)$ in MCE. The difference is just that the former includes the
Boltzmann weight $e^{-E/T}$, i.e. $\mathcal{Z}^{E}(V,T) ~\equiv~
Z^{MCE}(V,E)~e^{-\frac{E}{T}}$. Consequently the MCE multiplicity  
distribution is given by \cite{mce_massless,clt}:  
\begin{equation}\label{P_mce}
P_{MCE}(N) ~\equiv~ \frac{Z^{MCE}(V,E,N)}{Z^{MCE}(V,E)} ~=~
\frac{\mathcal{Z}^{N,E} (V,T)}{\mathcal{Z}^E (V,T)}~.
\end{equation}
Phrased differently, the $P_{MCE}(N)$ can be expressed as the conditional
distribution $P_{GCE}(N|E) \equiv P_{GCE}(N,E) /  P_{GCE}(E)$. 

The scaled variance of a multiplicity distribution is defined as:
\begin{equation}
\omega \equiv \frac{\langle N^2 \rangle - \langle N \rangle^2}{\langle N
  \rangle}~.
\end{equation}
It can be shown that in the large volume limit both distributions
(\ref{P_gce}) and (\ref{P_mce}) have the same mean value $\langle N \rangle$ ,
but rather different scaled variance \cite{mce_massless}. The scaled
variance in MCE converges to $\omega_{MCE} = 0.25$, while in GCE it is equal
to one, $\omega_{GCE} = 1$ (Poisson distribution). 

\section{Generalization to Multi-Specie Ideal Gas}
Above procedure can be easily generalized to a multi-specie hadron
resonance gas. When enforcing conservation of three Abelian charges, $Q^j =
(B,S,Q)$, as well as energy and three momentum, $E^k =
(E,P_x,P_y,P_z)$, the MCE partition function is given by \cite{MCEPF}:
\begin{eqnarray}\label{HRG_int}
\fl \mathcal{Z}^{Q^j,E^k} (V,T,\mu_j) =
    \Bigg[\prod \limits _{j=1}^3 \int \limits_{-\pi}^{\pi} \frac{d
      \phi_j}{2\pi} e^{-iQ^j \phi_j}\Bigg]
    \Bigg[\prod \limits _{k=1}^4 \int \limits_{-\infty}^{\infty} \frac{d
      \phi_k}{2\pi} e^{-iE^k  \phi_k}\Bigg]
     \!\exp \left[V \sum \limits_l \psi_l (\phi_j, \phi_k) \right],
\end{eqnarray}
where the single particle partition function is:
\begin{equation}
\psi_l (\phi_j, \phi_k)~=~ \frac{g_l}{\left( 2\pi\right)^3} \int d^3p
        \ln  \Bigg( 1\pm e^{-\frac{\sqrt{m_l^2+p^2}-\mu_l}{T} }~
        e^{iq^j_l \phi_j} e^{i\varepsilon_l^k   \phi_k } \Bigg)^{\pm 1},
\end{equation}
and the charge vectors $q^j_l$ and $\varepsilon_l^k$ of particle species
$l$, are defined as $q^j_l \equiv (b_l,s_l,q_l)$ and $\varepsilon_l^k \equiv
(\varepsilon_l,p_x,p_y,p_z)$. It can be shown that generally:
\begin{equation}
\mathcal{Z}^{Q^j,E^k}(V,T,\mu_j)~=~ Z^{MCE}(V,Q^j,E^k) ~e^{\frac{Q^j\mu_j}{T}}
        ~e^{-\frac{E}{T}}.
\end{equation}
Particle number `conservation` would result in an additional integral.

Above relation has two implications. The first one is technical. In MCE
calculations one has to deal with a heavily oscillating (or even 
irregular) integrand.  The integrants of Eqs.(\ref{massless_int},
\ref{HRG_int}) are however very smooth. For large volume the main  
contribution comes from small region around the origin, and an
analytical expansion is therefore possible \cite{SPE}.  $\mathcal{Z}^{Q^j,E^k}
(V,T,\mu_j)$ converges then to a  
Multivariate-Normal-Distribution, while finite volume corrections can be given
in the form of Hermite polynomials of low order \cite{clt}. The analytical
approximation is valid only around the maximum of the distribution.

The second implication is more subtle. Rather than calculating the
proper MCE partition functions, one can also, as in Eq.(\ref{P_mce}), define
$P_{MCE}(N)$ through a joint distribution of energy, momentum, charges and
particle number in GCE. $P_{MCE}(N)$ would then be called the conditional
probability distribution, while $P_{GCE}(N)$ is called the marginal
distribution of the partition function (\ref{HRG_int}). 

\section{Hadron-Resonance Gas Model and NA49 data}
In order to predict the energy depedence of $P(N)$ along the chemical
freeze-out line for Pb+Pb (Au+Au) collisions, the freeze-out parameters
should be expressed as a function of collision energy. Here we follow the
procedure described in \cite{mce,res}. The dependence of $\mu_B$ on the 
collision energy is parameterized as \cite{FOC}: $\mu_B \left( \sqrt{s_{NN}}
\right) =1.308~\mbox{GeV}\cdot(1+ 0.273~ \sqrt{s_{NN}})^{-1}~,$ where the
c.m. nucleon-nucleon collision energy, $\sqrt{s_{NN}}$, is taken in units of
GeV . The system is assumed to be net strangeness free, $S=0$, and to have the
charge to baryon ratio of the initial colliding nuclei, $Q/B = 0.4$. These two
conditions define the strange, $\mu_S$, and electric, $\mu _Q$,
chemical potentials. Temperature is defined by the condition: average energy
per particle is equal to $1~$GeV \cite{Cl-Red}. Finally, the
strangeness saturation factor, $\gamma_S$, is  parameterized as \cite{FOP} $
\gamma_S~ =~ 1 - 0.396~ \exp \left( - ~1.23~ T/\mu_B \right)$. This
determines all parameters of the model. Although various parameterizations are
known in the literature \cite{pbm,FOP}, resulting differences for multiplicity
fluctuations are small \cite{mce}. 

\begin{figure}
\includegraphics[width=7cm]{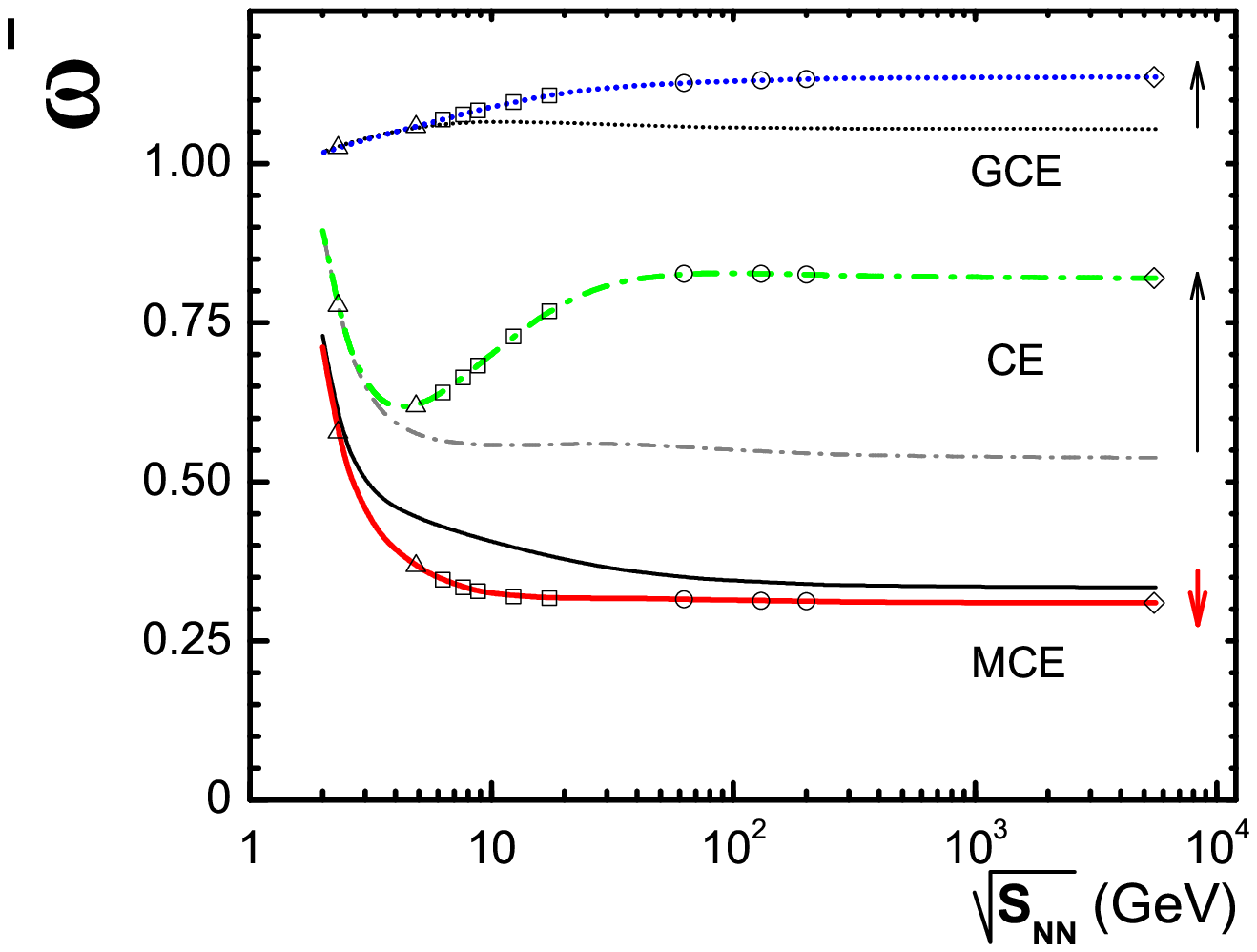}
\includegraphics[width=7cm]{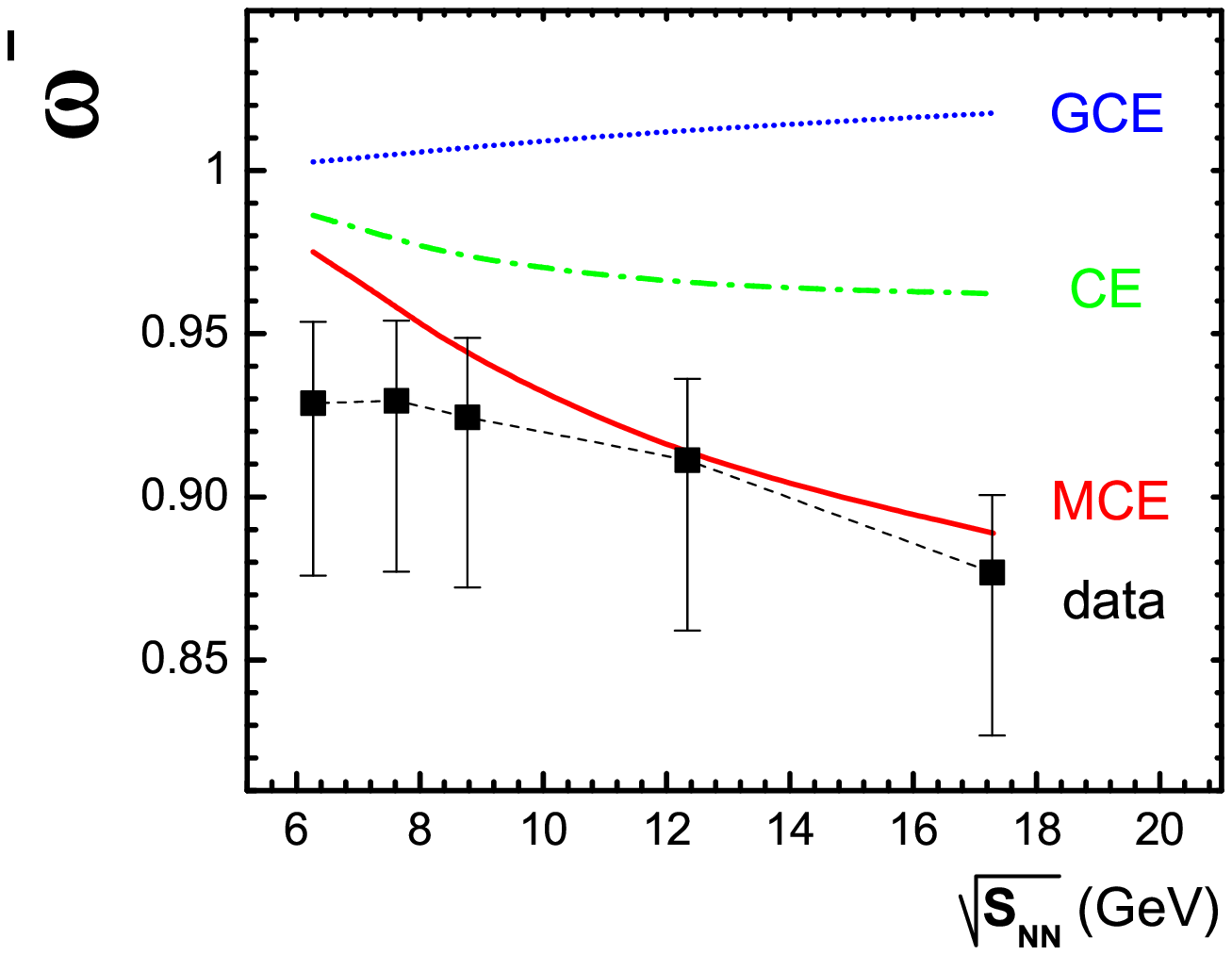}
\caption{Left: The scaled variances for negatively charged particles,
  $\omega^-$, both primordial and final, along the chemical freeze-out line
  for central Pb+Pb (Au+Au) collisions. Different lines present the GCE, CE,
  and MCE results. Symbols at the lines for final particles correspond to 
  specific collision energies ranging from SIS to LHC energies. The arrows
  show the effect of resonance decay. Right: The lines show acceptance
  corrected values for $\omega^-$ in the SPS energy range. The points show 
  preliminary data of NA49 \cite{NA49}. Total (statistical+systematic) errors
  are indicated. Figures are taken from \cite{mce}}\label{omega} 
\end{figure}

\begin{figure}
\includegraphics[width=14cm]{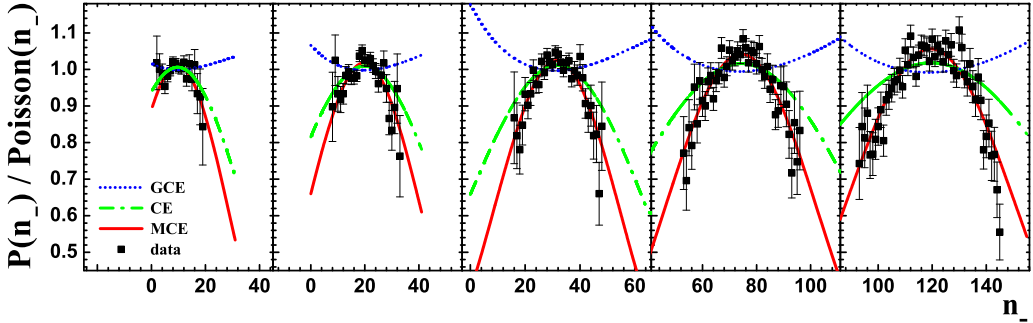}
\caption{The ratio of the multiplicity distributions to a Poisson reference
  distribution with same mean value for negatively charged hadrons produced in
  central (1\%) Pb+Pb collisions at 20$A$~GeV, 30$A$~GeV, 40$A$~GeV,
  80$A$~GeV, and 158$A$~GeV (from left to right) in the NA49 acceptance
  \cite{NA49}. Preliminary experimental data (solid points) of NA49
  \cite{NA49} are compared with prediction of the hadron-resonance gas model
  obtained within different statistical ensembles, the GCE (dotted lines), the
  CE (dashed-dotted lines), and the MCE (solid lines). Figures are taken from
  \cite{mce}}\label{Data}  
\end{figure}

For the calculations, an extended version of the THERMUS framework
\cite{Thermus} was used. The THERMUS particle table includes all known
hadrons and resonances up to a mass of about 2.5~GeV and their respective
decay channels. We use quantum statistics, but disregard Breit-Wigner width
of resonances. We assume that in the studied reactions the system volume is
large and finite volume corrections to fluctuations and mean values can be
neglected. Resonance decay has been dealt with as described in
\cite{clt,res}. The effect of finite experimental acceptance was taken into
account by an `uncorrelated particle` approximation \cite{CE,clt,res}. 
The results for multiplicity fluctuations of negatively charged hadrons in the
full momentum space and a comparison to data in the NA49 acceptance are shown
in Fig. \ref{omega}. The multiplicity distributions in the NA49 acceptance for
central $Pb+Pb$ collisions at $20$, $30$, $40$, $80$, and $158AGeV$ together
with the predictions for the GCE, CE, and MCE versions of the model are shown
in Fig. \ref{Data}.   

The measured multiplicity distributions are significantly narrower than the
Poisson one and allow to distinguish between model results derived within
different statistical ensembles. The data agree surprisingly well with the
expectations for the micro-canonical ensemble and exclude the canonical and
grand-canonical ensembles.  

\ack{ 
We would like to acknowledge frequent and fruitful discussions with
F. Becattini, E. Bratkovskaya, L. Ferroni, S. H\"aussler, and G. Torrieri.
MH would like to thank the organizers for financial support and for a
greatly enjoyable conference environment.
}

\end{document}